  \documentclass[12pt,preprint]{aastex}
 \usepackage{ifpdf}
 \usepackage{epsfig}
  \usepackage[latin1]{inputenc}
 \usepackage{amsfonts}
 \usepackage{amsthm}
 \usepackage{amssymb, latexsym, amsmath}
 \usepackage{amsbsy}
\usepackage{amstext}
\usepackage{amsopn}
\usepackage{amsgen}
\usepackage[dvips]{color}
 \usepackage{graphicx}
  \usepackage{color}
\newcommand{\eb}{\begin{equation}}
\newcommand{\ee}{\end{equation}}
\newcommand{\st}{KIC 7341653}
\shorttitle{Monster flares on KIC 7341653}
\shortauthors{Makarov \& Goldin}
\begin{document}
\title{Kepler data on KIC 7341653, a nearby M dwarf with monster flares and a phase-coherent variability}
\author{Valeri V. Makarov}
\email{valeri.makarov@navy.mil} 
\affil{US Naval Observatory, 3450 Massachusetts Ave NW, Washington DC 20392-5420, USA}
\author{Alexey Goldin}
\email{alexey.goldin@gmail.com} 
\affil{Teza Technology, 150 N Michigan Ave, Chicago IL 60601, USA}

\date{Accepted . Received ; in original form }

\label{firstpage}
\begin{abstract}
KIC 7341653 is one of several late-type M dwarfs observed in the main mission of Kepler with peculiar infrared colors placing them in the
domain of suspected young stellar objects (YSO). It is likely associated with a powerful X-ray emitter with X-ray
flares. Kepler light curves reveal two distinct types of activity: frequent flares lasting from less than 30 min to a few hours
and a periodic variability with a period of 0.5463441(7) d. The largest flare detected increased the flux in the Kepler passband by a factor
of 2.8 and released an estimated 4$\times10^{34}$ erg of energy in the Kepler band. 
Segmented periodogram analysis reveals that the amplitude of the periodic variation was subject to secular changes, dropping from
peak values around 20 ppt to below 5 ppt toward the end of the mission,
while the phase varied periodically with an amplitude of 0.15 radians and period 362(3) d. Two possible interpretations of the phase periodicity
are discussed: a migrating long-lived photospheric spot and 
a Doppler frequency shift generated by a solar-mass faint companion, such as a white dwarf. \end{abstract}


\section{Introduction}
The star \st~ drew our attention when we investigated a sample of Kepler main mission targets listed in the selection by \citet{mar}
of candidate class I/II/III young stellar objects (YSO), i.e., very young protostellar formations. The authors used near- and mid-infrared
photometric data from the 2MASS \citep{cut} and WISE \citep{cut13} catalogs and applied an automated method of machine learning.
We found 10 objects of class I/II and 175 objects of class III out of the $133,980$ YSO candidates matching Kepler targets with their positions 
within $10\arcsec$. These objects must be
interlopers with unusual infrared photometric properties mimicking those of YSO, since no known regions of ongoing star formation are present in the
field of view of Kepler. Even though young stars can be dispersed or ejected out of their parent clusters \citep[e.g.,][]{mak06},
the nearest ones are invariably found in loose associations or moving groups \citep{mak07}, because their ages (less than several
Myr) are too short for them to travel far. Our original motivation was to see if any extragalactic active galactic nuclei (AGN) could be identified
on the Kepler list, due to a partial overlap of YSO and AGN colors in WISE magnitudes \citep[see, e.g.,][]{sec}. Instead, we found a
versatile collection of stars, some of them quite blue in the optical such as a few A-type main sequence dwarfs, with variable light curves
and signs of activity, including small and frequent flares and dips. A smaller group of very red stars stands out with their
significant proper motions measured in the URAT1 catalog \citep{zac} and significant ground-based parallaxes \citep{fin}.
These stars are nearby late-type M dwarfs typically at 20--50 pc from the Sun. 

M dwarfs are very common stars in the Galaxy and in the Solar neighborhood in particular \citep{hen}, but the late-type objects found in the
intersection of the lists of YSO candidates and Kepler targets stand out from the general sample in a few respects. Firstly, their
unusual MIR colors resulted in the confusion with YSO and imply a significant presence of circumstellar dust. Whether the dust is aggregated
in accretion disks similar to the classical T Tauri stars or other types of structures, is not known. Secondly, the selected objects have other
signs of activity likening them with pre-main-sequence stars, such as flares and intense X-ray radiation. Many of these objects are
cross-matched with the Rosat Bright Source or Faint Source catalogs \citep{vog1, vog2}. A brief revision of the Kepler long-cadence light
curves reveals frequent flares (probably, of Sun-like, or chromospheric, type), sometimes accompanied with frequent dips of short duration
and periodic or quasi-periodic variations. Kepler data on a well-known representative of flaring M5 stars, the binary GJ 1245 A and B,
were studied in depth by \citet{lur}, including the total energies and effective duration of flares. 

\st~ has received much less attention in the literature. It is a very red star, with accurately determined AAVSO magnitudes $V=14.348(55)$ mag and
$B-V=1.549(65)$\footnote{Throughout this paper, numbers in round brackets stand for standard errors of values in the last
significant digits, i.e., $1.549(65)$ is equivalent to $1.549\pm 0.065$}
 mag \citep{he} placing it in the M4 spectral type domain. \citet{ria} estimate a photometric distance of only 14 pc and a relative X-ray luminosity
$\log L_X/L_{\rm bol}$ of -2.41, indicating a high degree of coronal activity surpassing the saturation limit at $\sim -3$.
\citet{fin} determined a larger trigonometric distance of 21 pc. Assuming this distance, the flux in the 0.3--10 keV band measured by SWIFT \citep{del}
converts to an X-ray luminosity of $3.14\times 10^{28}$ erg s$^{-1}$, which implies an even higher $\log L_X/L_{\rm bol}$ of -2.10.
The difference between the photometric and astrometric distance moduli of almost 1 mag may indicate the presence of a close, unresolved companion
which is fainter than the primary by 0.3 mag. This hypothetical companion could be a late-type M dwarf, but its presence would still not
explain the unusual infrared colors. \citet{ria} determined a rather low equivalent width of the H$_\alpha$ line, EW H$_\alpha =1.2$ \AA,
which is at the low end of the range for magnetically active stars according to \citet{wes}. This modest emission seems at odds with the
high level of X-ray luminosity.

\begin{figure}[htbp]
  \centering
  \includegraphics[angle=0,width=0.32\textwidth]{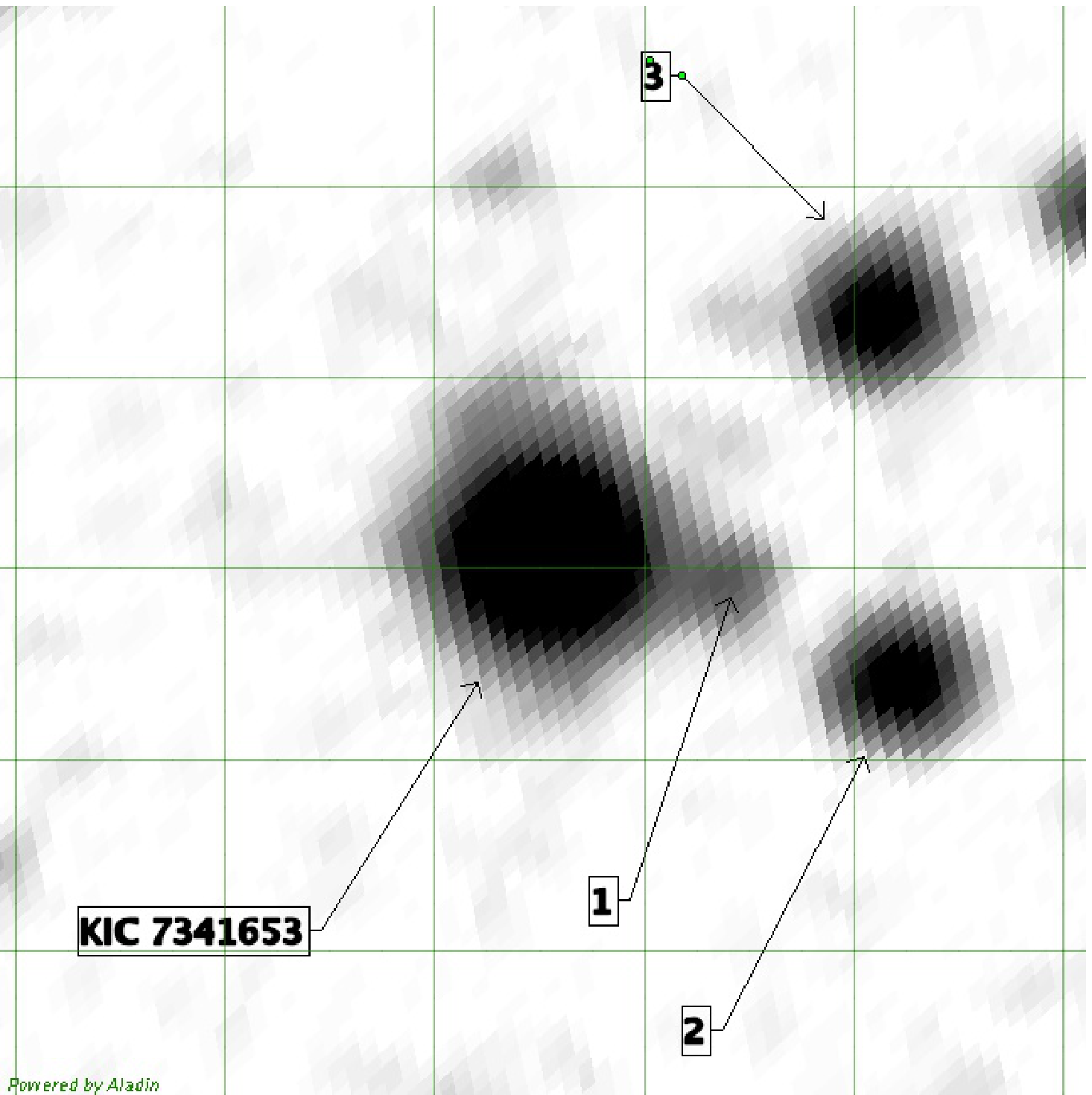}
  \includegraphics[angle=0,width=0.32\textwidth]{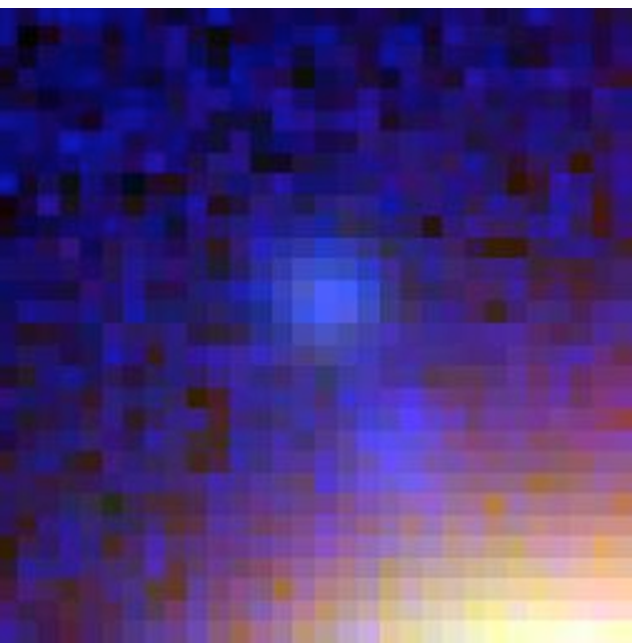}
  \includegraphics[angle=0,width=0.32\textwidth]{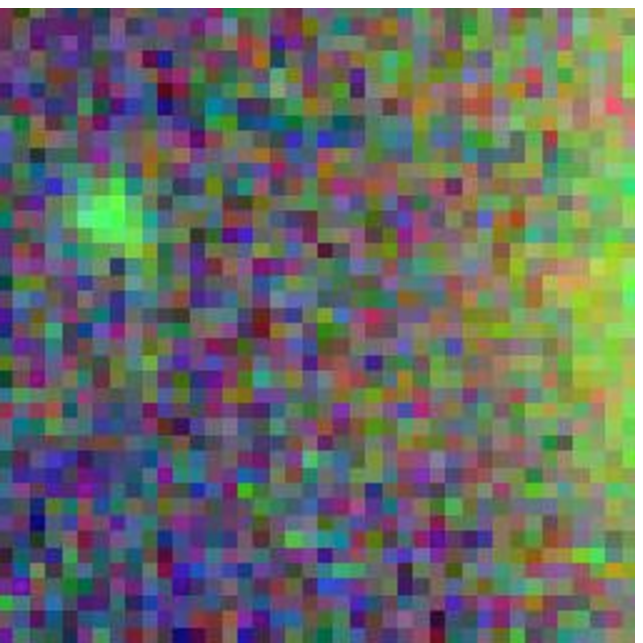}
\hspace{2pc}
\caption{Images of the sky at \st\ and its close surroundings. In all graphs, north is up, east is to the left. 
Left: Finding chart for  \st~ reproduced using the CDS Aladin v9.0 web tool from
the DSS Red survey. The size of a grid square is $10\arcsec$. The target star and three
neighboors discussed in the paper are labeled. Center: Stacked multi-color Pan-STARRS image of the close vicinity north of \st. 
The glow in the lower right corner indicates the
location of the star (outside the imaged area). The size of the image is $7\farcs5$ on a side. 
The source close to the center is the 20.7 mag Gaia source 2105033123254264320. The fainter blurred object between the source and the star
is not listed in any catalog. 
Right: Stacked multi-color Pan-STARRS images of the close vicinity east of \st, with the same scale as the center chart. The size of the image is
$10\arcsec$ on a side.
The peculiarly shaped faint object has not been seen or listed in any catalogs
or surveys.}
\label{map.fig}
\end{figure}

The associated X-ray source (1RXS J$185504.7+425952$) was detected by Rosat and included in the catalog of bright sources \citep{vog1}. Due to the poor
angular resolution of the Rosat telescope ($8\arcsec$ in this case), this association can not be a definite proof that the X-ray 
emission comes from the star
in question. Any source within several arcseconds of the star can in fact be responsible for this detection. \citet{fuh} identified \st~ as
a flaring X-ray source based on the Rosat all-sky survey, which implies a high degree of variability. The Gaia DR1 catalog \citep{bra} lists our target
at (RA, Dec)J2015 = (283.7690559, +42.9975114) deg, and with $G=12.633$ mag, but there is a much fainter neighbor in that catalog, source 2105033123254264320,
at (283.7700490, +42.9995093), $G=20.656$ mag, i.e., separated by $7.653\arcsec$ in PA=$20.0\deg$ from our target. The Rosat position, for what it's worth,
is closer to the bright star's position, but in the catalog of optical/X-ray associations \citep{fle}, the only source listed (MORX J185504.7+425957) is much closer to
the faint Gaia neighbor, and is associated with it. The optical counterpart is considered to most likely be a galaxy. This would seal the fate of
this X-ray association perhaps, if not for the more accurate position of the X-ray detection with Swift \citep{eva} at (283.76943, +42.99797) deg with a 90\%
radial error of $4.4\arcsec$, which is closer to the M dwarf. We should accept that there is no complete certainty in where the large and variable
X-ray emission comes from. Fig. \ref{map.fig}, center, shows the stacked Pan-STARRS DR1 image centered on the Gaia position of the faint neighbor. A star-like
source is clearly seen close to the expected position, but there is another object between it and the target star whose glow is visible in the lower right
corner. This faint object, possibly a galaxy, is separated from the M dwarf by approximately $5\farcs6 $ in PA=20$\degr$ is another candidate for the
bright X-ray emission.

The brightest X-ray emitters in the immediate solar neighborhood are predominantly RS CVn-type giants, BY Dra-type main sequence dwarfs, short-period
binaries of W UMa type, and pre-main-sequence or T Tauri stars \citep{mak03}. The excess of coronal X-ray emission for these normal field
stars is confidently related to a fast rotation caused by either tidal interaction in binaries or a young age. M dwarfs are relatively fainter in
X-rays despite the existence of the population of very active, emission-line dMe stars. Even when such magnetically active M dwarfs reach the
saturation limit, they can not rival their larger counterparts in the total X-ray coronal emission due to the smaller volume of their coronas. The
possible (but not confidently proven) high X-ray luminosity of \st~ singles it out as a very unusual object. Our original motivation was to
find out if Kepler data bore any evidence of close binarity or rapid rotation. To this end, we processed and analyzed the entire sample
of Kepler astrometric and photometric data available through the MAST archive hosted at the STScI, covering almost 4 years of continuous
observations in the long-cadence regime. The methods of data processing are briefly described in Section \ref{met.sec}. Section \ref{fla.sec}
deals with photometric flares detected in the PCA-processed light curves. Variability-induced motion (VIM) effects in photometry and astrometry are
presented in Section \ref{vim.sec}. Periodic variations of the light curves are analyzed in Section \ref{perio.sec}. Interpretation of the
results is given in Section \ref{int.sec}, and a summary in Section \ref{sum.sec}.

\section{Methods}
\label{met.sec}
We use the raw Simple Aperture Photometry (SAP) Kepler light curves rather than the Pre-search Data Conditioning (PDC)
fluxes derived by the data processing pipeline \citep{chr}. Our aim was to detect possible variations in the data on
the time scale up to 90 days. The de-trending procedures based on low-order polynomials
to remove the slow variations of flux in the raw data are not always sufficient to mitigate the common instrumental
perturbations shared by targets on the same channel. Instead we applied our specially designed Principal Component
Analysis (PCA) pre-whitening and filtering procedure decribed in \citet{mak16}.
This method allows us to come as close as possible to
the intrinsic variability of the star at a level of 20 ppm. We always use 4 principal components for both astrometric and
photometric data divided into quarter-year segments ($\sim90$ days).

We add to our analysis the astrometric data gathered by Kepler, i.e., the photocenter centroids
measured at the same cadence and simultaneously with the flux. Synchronous astrometric excursions of centroids, called Variability
Induced Motion (VIM),
provide a powerful method of detecting double stars or other events of signal blending \citep{mak}. 
VIM occurs when one of the blended sources in the aperture is
photometrically variable. Since the recorded centroid corresponds to the first moment of the distribution of
flux over the pixel grid, which is sensitive to the flux ratio of the blended components, the photocenter moves
in a strongly correlated way with the total flux. The published list of VIMs includes 12 confident detections (for
12 mission quarters) with a rather large effective distance of $ds/dF\times F50\simeq 0\farcs2$, and VIM directions
scattered around $75\degr$ on the sky. 

\section{Monster flares}
\label{fla.sec}
\begin{figure}[htbp]
\epsscale{1.1}
\plottwo{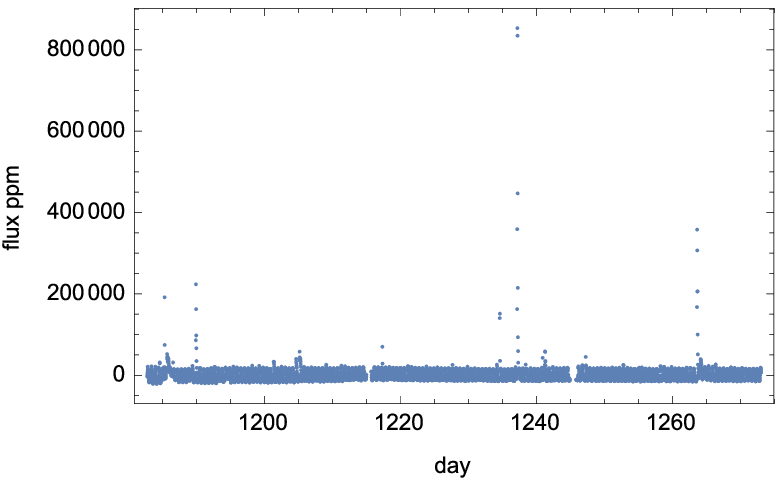}{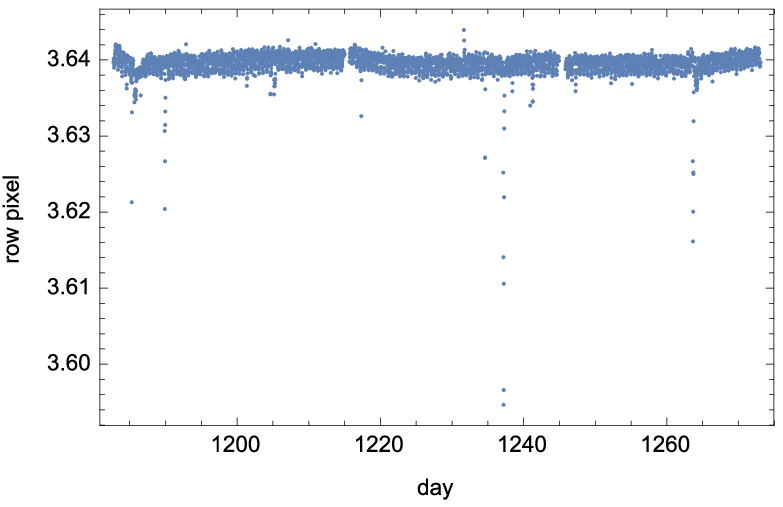}
\hspace{2pc}
\caption{PCA-cleaned normalized SAP flux (left) and centroid position in row pixels (right) of \st~ during mission quarter 13. Multiple flares
were observed, mirrored in the centroid position due to a VIM effect.}
\label{q13.fig}
\end{figure}

Fig. \ref{q13.fig}, left, shows the PCA-filtered 30 min-cadence light curve in Quarter 13 normalized to the median value. The median detected flux is
$\sim 93\times10^3$ e$^-$ s$^{-1}$ , which puts the object below the blooming and saturation limits. The standard deviation of the flux values is 
$\sim3900$ e$^-$ s$^{-1}$.
Multiple spikes of variable height indicate flare activity on this star. This does not come as a surprise because mid-M dwarfs are known to
have energetic and frequent flares even when their H$\alpha$ measurements do not indicate an elevated magnetic activity \citep{haw}. Probably the
best studied flaring M dwarfs in the Kepler collection are GJ 1243 (M4) and GJ 1245 AB (M5). The former is especially close to \st~ by its
general characteristics, although it has a higher EW H$\alpha$. But the flares we find on \st~ are of immense power compared to its previously
known counterparts, including the famous prototype of the class, UV Ceti \citep{lov}, and rivaling the BY Dra-type system YY Gem as presented
in \citet{lac}.
 We call them ``monster flares" in analogy with the ``super-flares" recently suggested for solar-type stars. Table ~1 lists
the most prominent flares detected in the set of 16 mission quarters with amplitudes confidently exceeding 10\% (100 parts-per-thousand, ppt) relative to the quiescent flux.
Peak fluxes can only approximately be estimated with the available light curves because of the insufficient time resolution of the 30-min cadence.
Likewise, the duration and the total emitted power are roughly estimated. The latter parameter, $E_{Kp}$ in units of erg, is derived following the
procedure described in \citet{haw}. Using their specific flux zero-point and the trigonometric parallax, we estimate the quiescent luminosity 
of \st\ in
the Kepler band at $\log L_{Kp}=30.70$ erg s$^{-1}$. This is close to the previously estimated $\log L_{Kp}$ for GJ 1243 (30.67). Column 1
specifies the time of peak flux for each flare in mission days, column 2 the peak amplitudes in ppt, column 3 the total energy, $\log E_{Kp}$, in
erg, column 4 approximate total duration in hours, and column 5 notes about the structure and associated astrometric signals.

 \begin{table*}
 \centering
 \caption{Largest flares of \st.}
 \label{fla.tab}
 \begin{tabular}{@{}lrcll@{}}
 \hline
            &                 &       & &\\
   Day     &  Amplitude    & $\log E_{Kp}$ & Duration  & Notes\\
           &  ppt          &    erg        &    hr     & \\
 \hline
 191.1 & 194 & 33.7  &  3.0 &\\
 216.4 & 183 & 34.2  & 12.0 & complex \\
 275.0 & 124 & 33.2  & 1.5  & flare and wave have diff. VIM direction \\
 369.4 & 187 & 33.3  & 2.0  & \\
 377.6 & 155 & 33.5  & 1.5  & \\
 389.1 & 158 & 33.5  & 4.5  & \\
 409.6 & 140 & 33.5  & 2.5  & \\
 523.1 & 204 & 33.5  & 2.0  &\\
 551.9 & 177 & 33.6  & 2.0  & \\
 560.2 & 134 & 33.4  & 3.0  &\\
 602.1 & 133 & 33.5  & 2.5  & \\
 659.3 & 570 & 34.0  & 2.0  & \\
 668.4 & 508 & 33.9  & 2.0  & \\
 675.4 & 314 & 34.4  & 10.0 & complex\\
 695.6 & 349 & 34.1  & 4.0  & \\
 863.5 & 204 & 33.3  & 2.5  & \\
 997.5 & 630 & 34.6  & 10.0 & double, sep. by 3 hr\\
1117.5 & 133 & 33.4  & 2.0  & \\
1237.2 & 834 & 34.4  & 5.0  & \\
1263.7 & 358 & 34.1  & 5.0  & may be double\\
1319.5 & 234 & 34.2  & 7.0  & complex\\
1339.8 & 137 & 33.3  & 2.0  & \\
1385.8 & 106 & 33.0  & 1.5  & \\
1391.9 & 139 & 33.3  & 2.0   & \\
1426.4 & 185 & 33.5  & 2.0  & \\
1438.5 & 142 & 33.7  & 5.0  & double,sep. 2.5 hr, diff. VIM direction\\
1443.1 & 1773 & 34.6 & 3.5  & most powerful flare detected\\
1452.3 & 214 & 33.5  & 2.5  & \\
1464.6 & 338 & 33.9  & 6.0   & diff. VIM direction\\
1577.1 & 104 & 33.3  & 2.0  & \\
1579.0 & 253 & 33.8  & 2.5  & \\
 \hline
 \label{table}
 \end{tabular}
 \end{table*}

It is quite remarkable how much energy is released in the flare events on \st. These are roughly 30 times more powerful than the
flares on the close analog GJ 1243. The largest detected flare on day 1443.1 was so powerful that the flux peaked at 2.8 times
the quiescent level. A burst of similar magnitude, if it happened on the Sun, could be catastrophic for the ecosphere on Earth. 
However, even the greatest ``super-flares" observed on solar-type stars \citep{mae} fall far behind the energy we observe for \st.

\begin{figure}[htbp]
\epsscale{1.1}
\plottwo{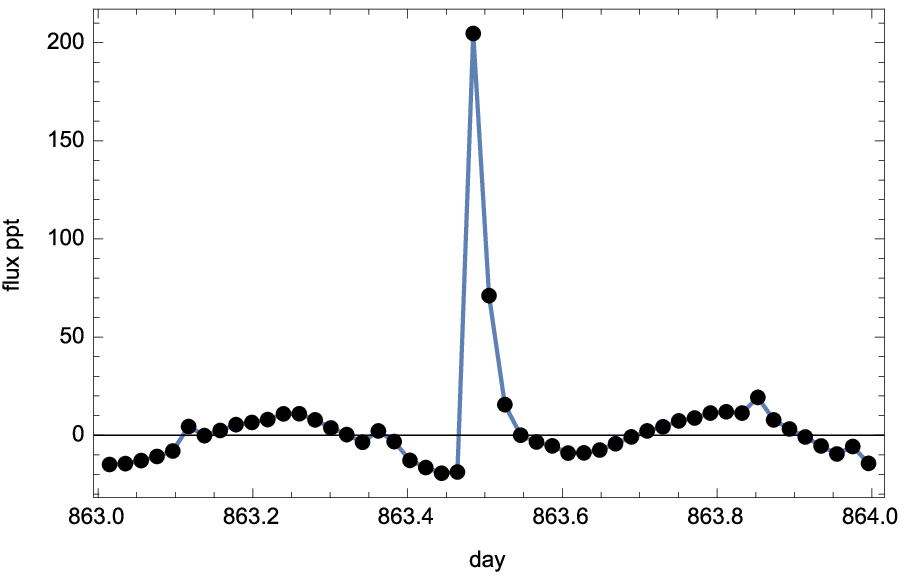}{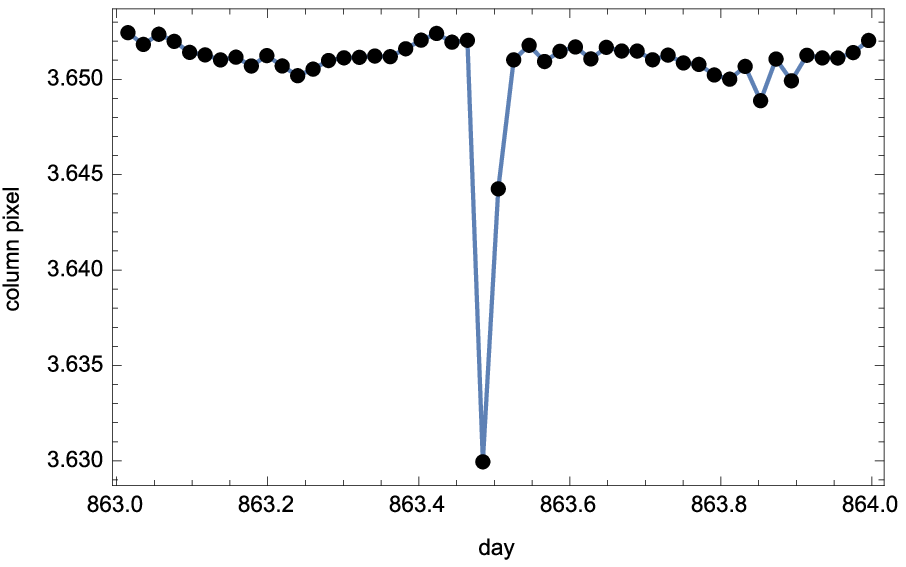}
\hspace{2pc}
\caption{PCA-cleaned normalized SAP flux (left) and centroid position in column pixels (right) of \st~ during the flare event
on day 863.5.}
\label{singflare.fig}
\end{figure}

The morphology of these powerful flares is diverse. Fig~\ref{singflare.fig}, left, depicts a single, short duration flare on day 863.5, characterized by
a rapid rise, often not resolved on the 30-min cadence, and a slower exponential decay lasting for one to a few hours. It appears that this sharp and powerful burst in flux is faithfully reproduced in the column pixel coordinate
shown in the right panel of the figure. Quite a few
flares are distinctly double, with time-resolved peaks, or more complex multiple flares with sometimes gradual rising slopes, such as
the one on day 675.4. Only a small fraction ($\sim10$\%) of flare events have such complex structures, possibly indicating a different
physical origin.

\section{VIM detections}
\label{vim.sec}

The {\it en masse} analysis by \citet{mak} has already established that \st~ shows powerful and consistent VIM effects during the
entire mission. The prevailing direction of the astrometric perturbations (position angle counted from north through east)
is $\sim 75\degr$. We find a previously unknown,
very faint and possibly elongated object approximately in this direction with J2000 coordinates $(283.7755,\,+42.998)$,
separated by $10\farcs3$ at position angle $82\degr$ (Fig.~\ref{map.fig}, right). This object has not been listed in any catalog or survey. It can be discerned only
in heavily stacked Pan-STARRS images, implying a faint magnitude. This enigmatic object can hardly cause the observed centroid
motion due to its faintness. The close blue companions in Fig. \ref{map.fig}, left, have a position angle inconsistent with
the detected VIM direction. Recalling that the VIM vector is inverted when the photometric variability takes place on the
target star (i.e., a flare on the target star will move the photocenter toward the target star and away from the blended companion),
we should consider the opposite direction to the average VIM vector ($\sim 255\degr$). We find another, much brighter, neighbor
with a Gaia DR1 position $(283.76517,\,+42.99702)$, separation $10\farcs38$ at position angle $260.2\degr$, and $G=19.29$ mag
(component 1 in Fig.~\ref{map.fig}, left).
The Pan-STARRS DR1 (PS1) PSF magnitudes for this star are $g=20.763(20)$, $r=19.584(11)$, $i=18.845(6)$, $z=18.481(11)$, $y=18.024(11)$
mag, indicating a background early M-dwarf. It is tempting to assign the observed VIM to this companion. A simple calculation of the
expected photocenter shift seems to be inconsistent with this interpretation, however. 
As the companion is approximately 500 times fainter than the target star,
the moment-based centroid should be displaced by roughly $0\farcs02$ in the quiescent state. No matter how much extra flux comes 
from a flare on the primary,
the astrometric excursion can not exceed this distance. The observed excursions during the largest flares exceed $0\farcs15$.

\begin{figure}[htbp]
\epsscale{1.1}
\plottwo{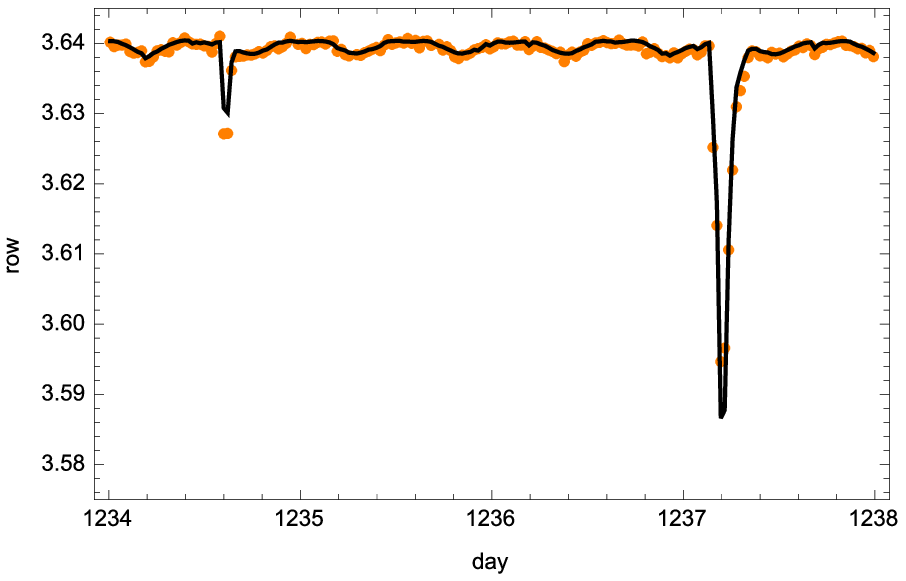}{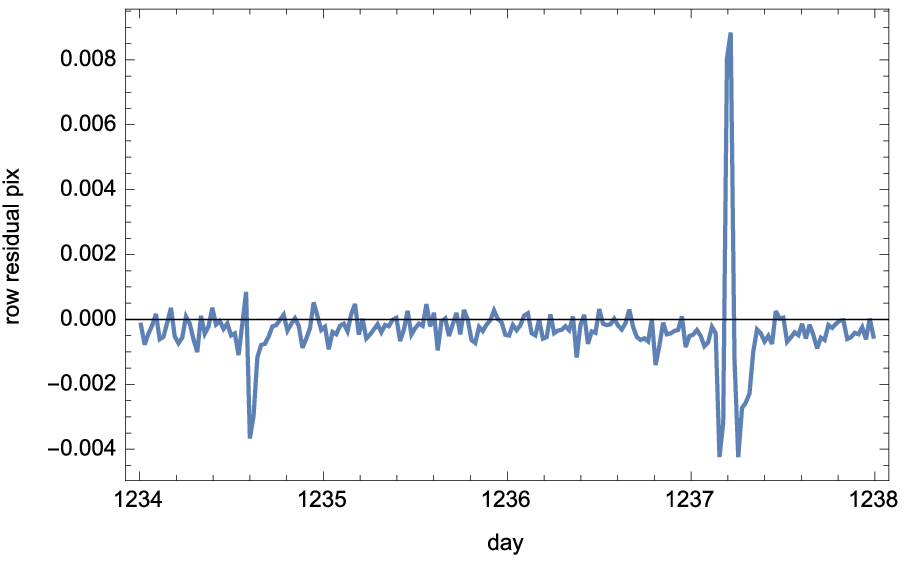}
\hspace{2pc}
\caption{Left: PCA-cleaned row pixel coordinates for a segment of long-cadence data including two flares (orange dots) and a
least-squares linear fit of the simultaneous light curve (black line, see text). Right: residuals of row centroid measurements
after the best fitting linear model based on flux is subtracted.}
\label{row.fig}
\end{figure}

\begin{figure}[htbp]
\epsscale{1.1}
\plottwo{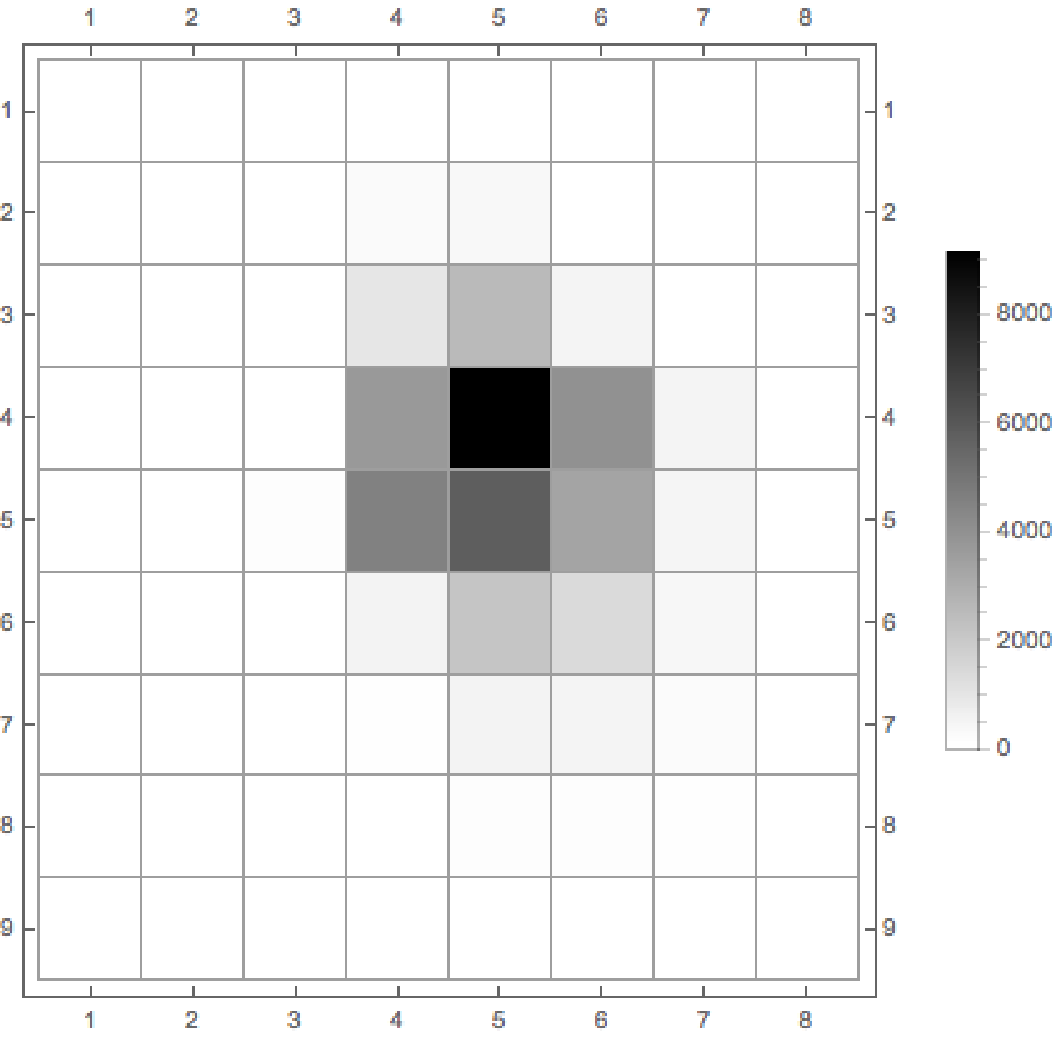}{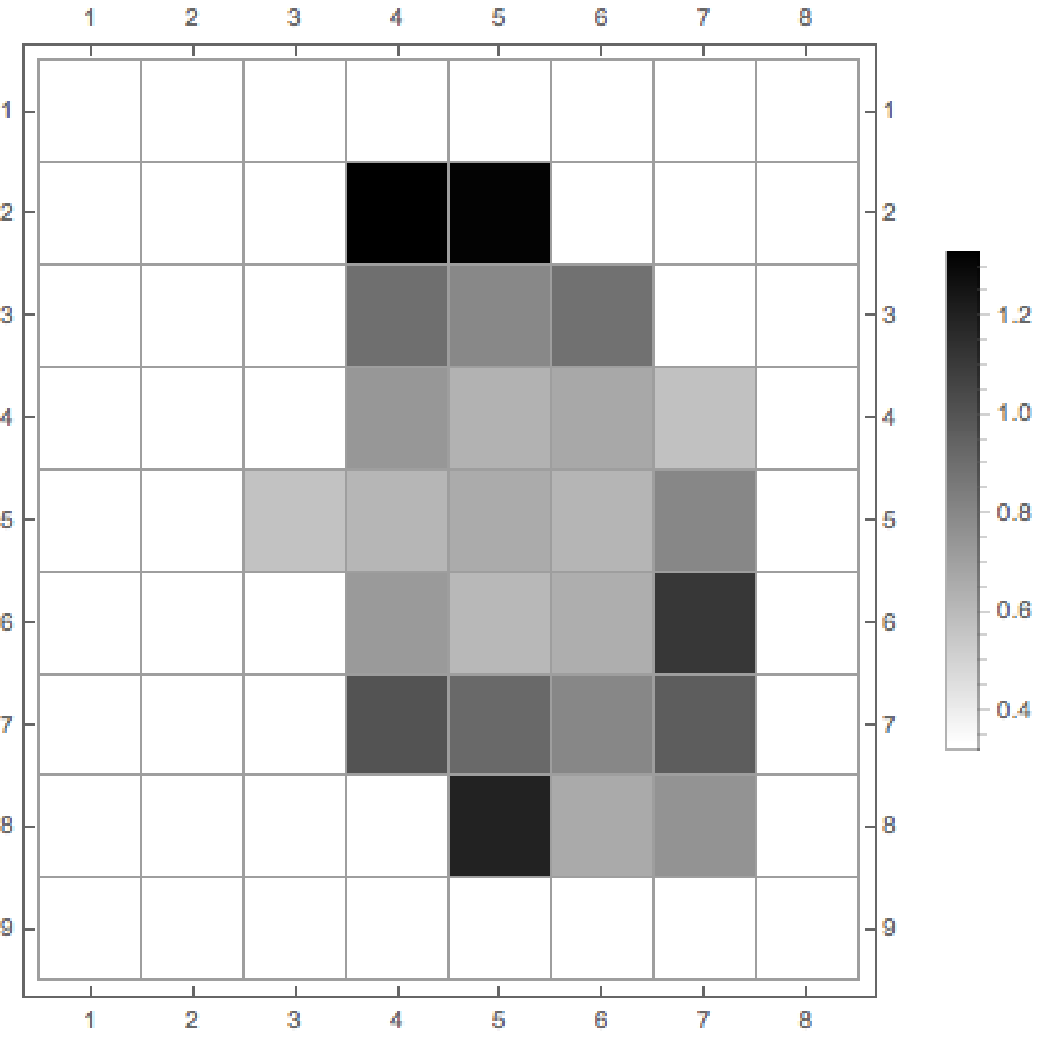}
\hspace{2pc}
\caption{Left: Intensity-coded pixel flux registered during the flare on day 1237.2 in excess of the median pixel values
over that quarter. The units are e s$^{-1}$.
Right: the same flare flux but relative to the median flux distribution. The shaded area shows the star- and
quarter-specific digital aperture used both for astrometry and photometry.}
\label{pix.fig}
\end{figure}

A much brighter, but also much more distant, neighbor labeled 2 in Fig. \ref{map.fig}, left, has also roughly the matching position angle
($250.0\degr$). It is separated from the target star by $20\farcs7$ and has a $G$ magnitude of 17.05. Being $\sim 0.017$ in flux of the target,
this companion could generate a photocenter shift of $0\farcs35$, if it were completely blended. The digital aperture is large enough
(Fig. \ref{pix.fig}) to accommodate part of the flux from neighbor 2, which can cause a VIM of the observed magnitude because of the
asymmetric distribution of light in the truncated image. The elongated shape of the aperture, which varied for the mission quarters,
could also explain why this companion, but not the companion 3 of similar brightness at separation $21\farcs6$, position angle $303.2\degr$
(neighbor 3 in Fig. \ref{map.fig}) generates the strongest VIM.

Having identified a plausible source of blended flux, we would like to make sure that the powerful flares detected in the light curve
do not come from this chance companion. Firstly, we note that the brightest flare on day 1443 raised the flux by a factor of 2.8, which
would imply an increase by 5.5 magnitudes if it comes from the companion. Although flares of such magnitude have been recorded for most
active binaries of BY Dra and RS CVn type, companion 2 does not seem to be a plausible source with its Sun-like colors ($g_{\rm PS1}-r_{\rm PS1}
= 0.44$ mag). Secondly, flare images do not show any obvious asymmetry one should expect from a source separated by $\sim5$ pixels from the target.
Fig. \ref{pix.fig}, left, shows the differential distribution of pixel flux during the peak of day 1237.2 flare event in absolute units of
electrons per second, i.e., the difference in flux between the maximum of that flare and the median flux over the entire Q13 data, for each pixel.
The center of the stellar image is in pixel $(5,4)$. The distribution of additional flux seems to be consistent with the quiescent flux distribution
and centered on the target star.

The same data expressed in relative units, i.e., $(f_{\rm flare}-f_{\rm quiescent})/f_{\rm quiescent}$, tells a different story (Fig. \ref{pix.fig},
right). While the total flux increased by 83\% during this flare (Table 1), some pixels on the fringes became brighter by
a larger fraction, and this enhancement is observed on both sides of the image. This implies that the shape of the image changed during the
flare, not just the amplitude. The two relatively brightest pixels, $(4,2)$ and $(5,2)$, defined the VIM direction for this event, but the
enhancement of flux in the counter direction is fairly obvious too. Such bizarre structures may occur, for example, if the subtracted
sky background or dark current value are overestimated, artificially reducing the level of starlight in pixels. The VIM direction is expected
to be resilient to this error, but it may alter the VIM magnitude by putting additional flux outside the blended companion. Another possibility
is a nonlinear response of pixels to increasing incident flux due to saturation. Saturation effects in Kepler data have been encountered and
discussed by \citet{lur, lug}. Our target star is significantly fainter than the $Kp=11.3$ mag which is believed to be the saturation
threshold \citep{gil}. However, scientific CCDs are known to have an interval of nonlinear quantum response before the full well saturation
is achieved \citep{jan}. It is possible that the most energetic flares drive the brightest pixels into the pre-saturation nonlinear
regime. The full well of the Kepler CCD is approximately $10\,000$ DN \citep{kib}, and at a typical inverse gain of 110 e$^-$/DN,
the corresponding pixel flux is $\sim180\,000$ e$^-$ s$^{-1}$. We can find some support to this hypothesis in Fig. \ref{row.fig} which shows the observed light curve (black line) fitted to the
observed centroid position in row coordinates (orange points) for a segment with two flares. While the fitted flux undershoots the centroid
motion for the small flare on day 1234.6, it overshoots the shift during the peak of the large flare on day 1237.2. This can also be seen in the
residuals of the fit in the right plot of Fig. \ref{row.fig}. This systematic behavior is found for the entire set of flares. It is a worrying
result, because it implies an onset of nonlinear response somewhere between $23\,000$ and $30\,000$ e$^-$ s$^{-1}$, which is a much lower value
than previously assumed. 

\begin{figure}[htbp]
\epsscale{1.1}
\plotone{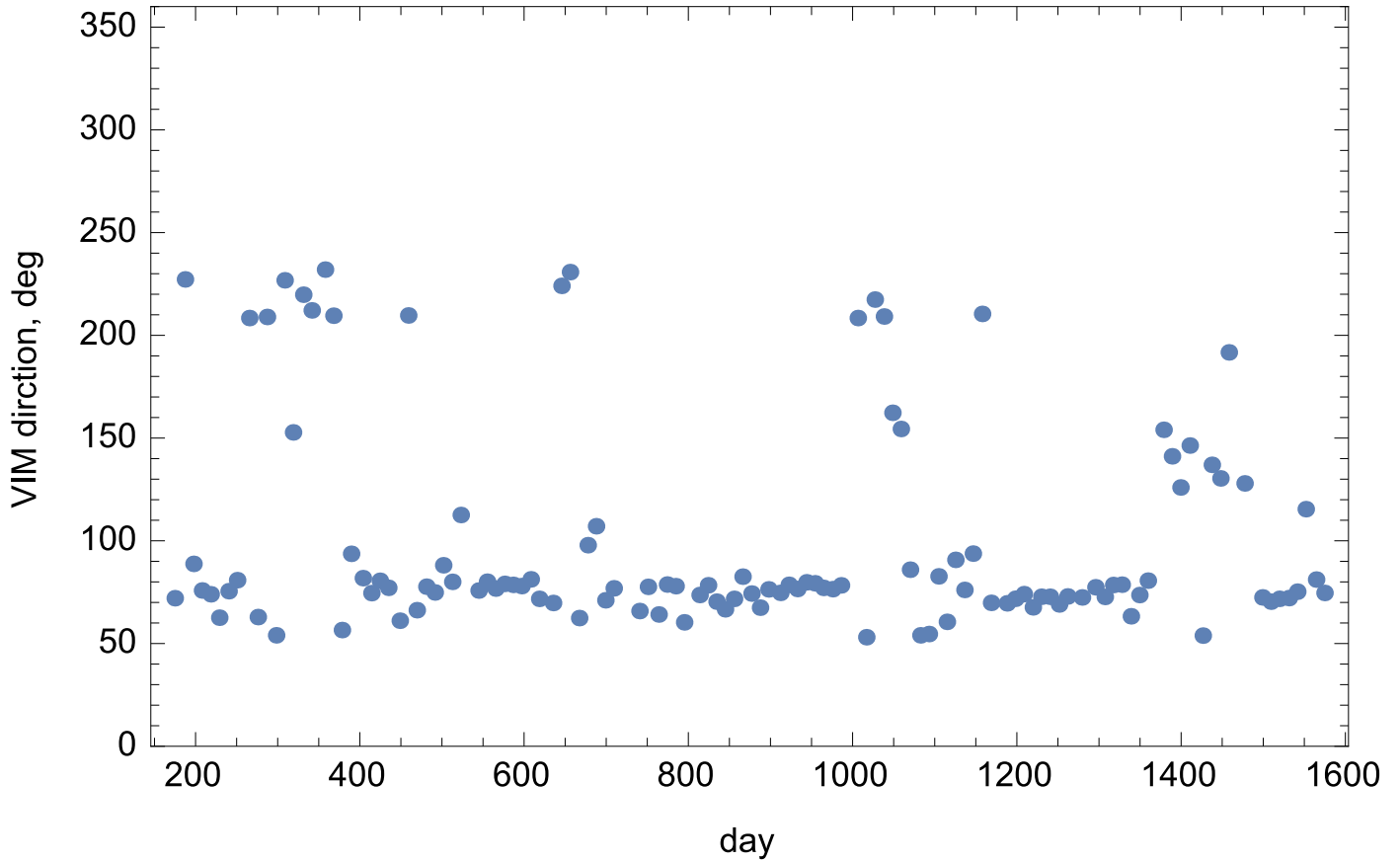}
\hspace{2pc}
\caption{VIM directions in sky position angles calculated for bins of 500 consecutive measurements (flux and centroid coordinates)
over the entire collection of 16 mission quarters.}
\label{vimpa.fig}
\end{figure}

We suggest an alternative explanation to the aureole of fractional flux generated by a bright flare. Pixels of a CCD do not have physical
boundaries but are delineated by so-called depletion zones where the quantum efficiency drops to zero. As charge is accumulated in a pixel,
the depletion zones shrink \citep{pla}. The boundary between two adjacent pixels will shift toward the pixel with more accumulated charge,
effectively reducing its collecting area. A fraction of photons will therefore be re-distributed between the pixels, with the fainter one
collecting relatively more light. This phenomenon is dubbed the ``brighter--fatter" (BF) effect \citep{ant}.
It is not visible in regular flat-field images of uniform intensity, but can be quite large for small-scale astronomical images such as
point-like sources \citep{smi}. E.g., the fractional difference between a long and a short exposures of the same source (light spot)
shows a lower peak and broader wings of the bright spot, which generate a doughnut-shaped distribution of light \citep{guy}. If this
interpretation holds, the BF aureole for Kepler star images is very wide, extending to $\sim5$--6 pixels in diameter. Due to the elongated
shape of the digital aperture, some fraction of the redistributed extra charge may fall outside the collecting area. The photometric effect
will be limited to the spillage of charge at the wings of the image, causing a small decrease of the total signal. Astrometric effects are
more significant, changing the magnitude of observed motion to higher values. As a conclusion, the so-called VIM speed parameter
may be exaggerated by the BF effect and carry little information about the configuration of blended sources.

With this caveat in mind, we consider in greater detail the direction of the VIM motion. Fig. \ref{vimpa.fig} shows VIM position angle
determinations for the whole long-cadence data set covering 4 years. Each point corresponds to a bin of 500 non-overlapping observations.
The estimation was done differently from the previous analysis by \citet{mak}. The coordinate measurements in $x$ (column) and $y$ (row)
for each bin were fitted with a straight line using the orthogonal regression method \citep{mak16}. The slope of this line defines the
coordinate direction which is converted into a sky position angle using the WCS parameters in the metadata. Most of the short-term estimates
cluster around a PA$=75\degr$, in a good agreement with the previous quarterly estimation by a different method. But there are occasional
intervals where the VIM direction switches to a much different value, which is $\sim 220\degr$ in the first half of the mission gradually
converging with the stable direction toward its end. This bifurcation can be explained by the presence of a third blended component with
an azimuthal motion around the target star. We thus find a clue to the existence of an orbiting companion with a period of longer than 4 years.
The magnitude of VIM depends on the amplitude of photometric variability, and if the hypothetical companion is variable, it can sometimes dominate the observed motion despite the smaller separation.

\section{Periodic variation of flux}
\label{perio.sec}
\begin{figure}[htbp]
\epsscale{1.0}
\plotone{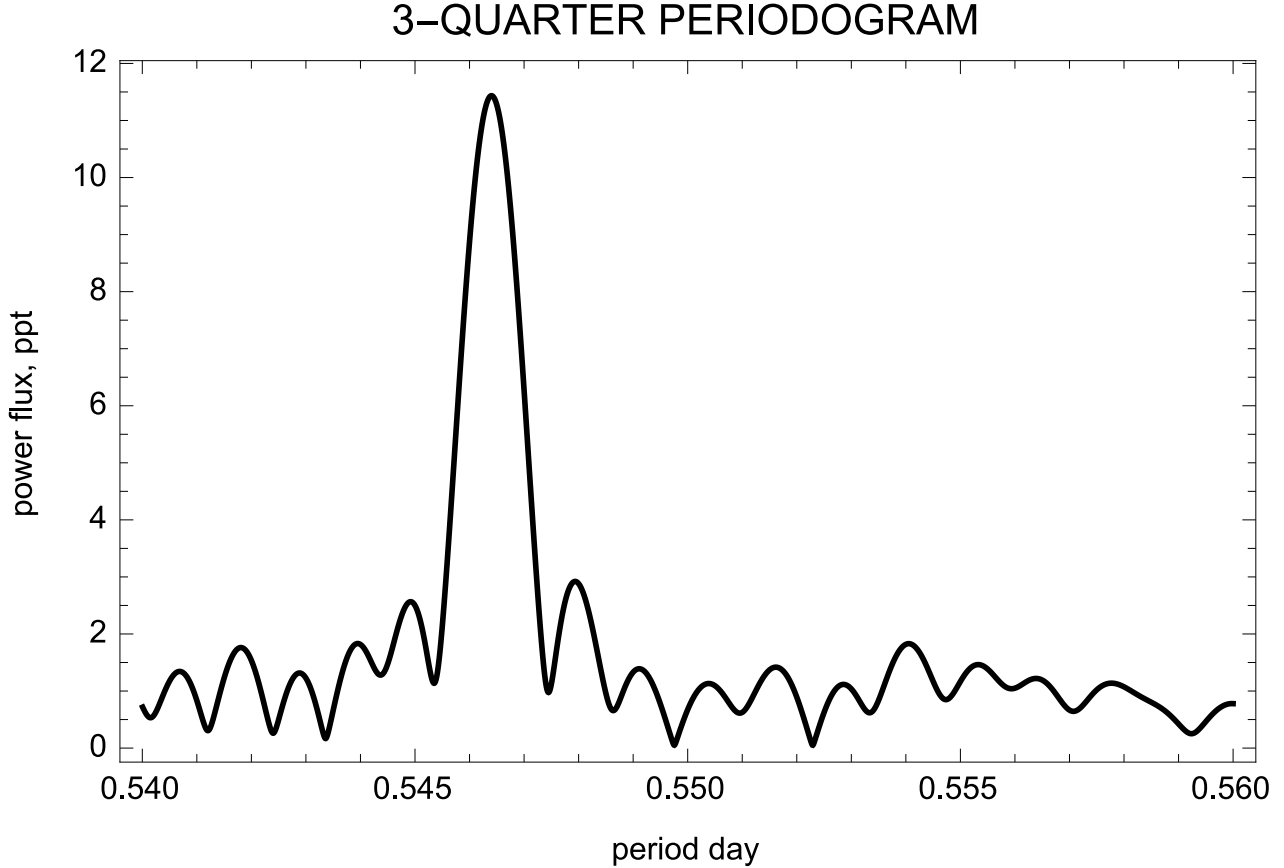}
\hspace{2pc}
\caption{High-resolution Lomb's periodogram of the joined light curve for 3 consecutive mission quarters, Q13 -- Q15. The dominating
peak at period 0.5464 d remains unresolved, and its width is roughly consistent with the windowing function.}
\label{3q.fig}
\end{figure}

We performed multiple periodogram computations parsing the PCA-cleaned light curve into segments of various length, as well as the entire mission.
The most energetic flares were clipped for a better characterization of the periodic wave (visible in Fig. \ref{row.fig}, for example). 
A powerful peak was invariably
obtained at a period $\sim0.5464$ d, with smaller secondary and tertiary harmonics clearly present, indicating a non-sinusoidal oscillation. Unlike
KOI-54, the prototype of the Heartbeat stars \citep{wel}, where the main peak splits in two for a sufficiently long interval of
data, our dominating signal remains unresolved in frequency in all the periodograms. Fig. \ref{3q.fig} depicts a regular unweighted
Lomb's periodogram of the oscillating light curve for three quarters of the mission combined. The peak with an amplitude of 12 ppt
remains single and unresolved suggesting a single-frequency periodic modulation. This separates the star in question from the
heartbeat class with tidally excited pulsation and high-eccentricity binary companions. On the other hand, the phase seems to be 
remarkably stable and coherent during the
entire 4 years of observation. A long-term coherent phase is usually associated with binarity and orbital motion. In analogy with the
eclipse time variation (ETV) method which reveals dynamical perturbation from tertiary companions of eclipsing binaries \citep{bor},
the phase of a periodic flux signal can be estimated for consecutive segments of the light curve. 

\begin{figure}[htbp]
\epsscale{1.0}
\plotone{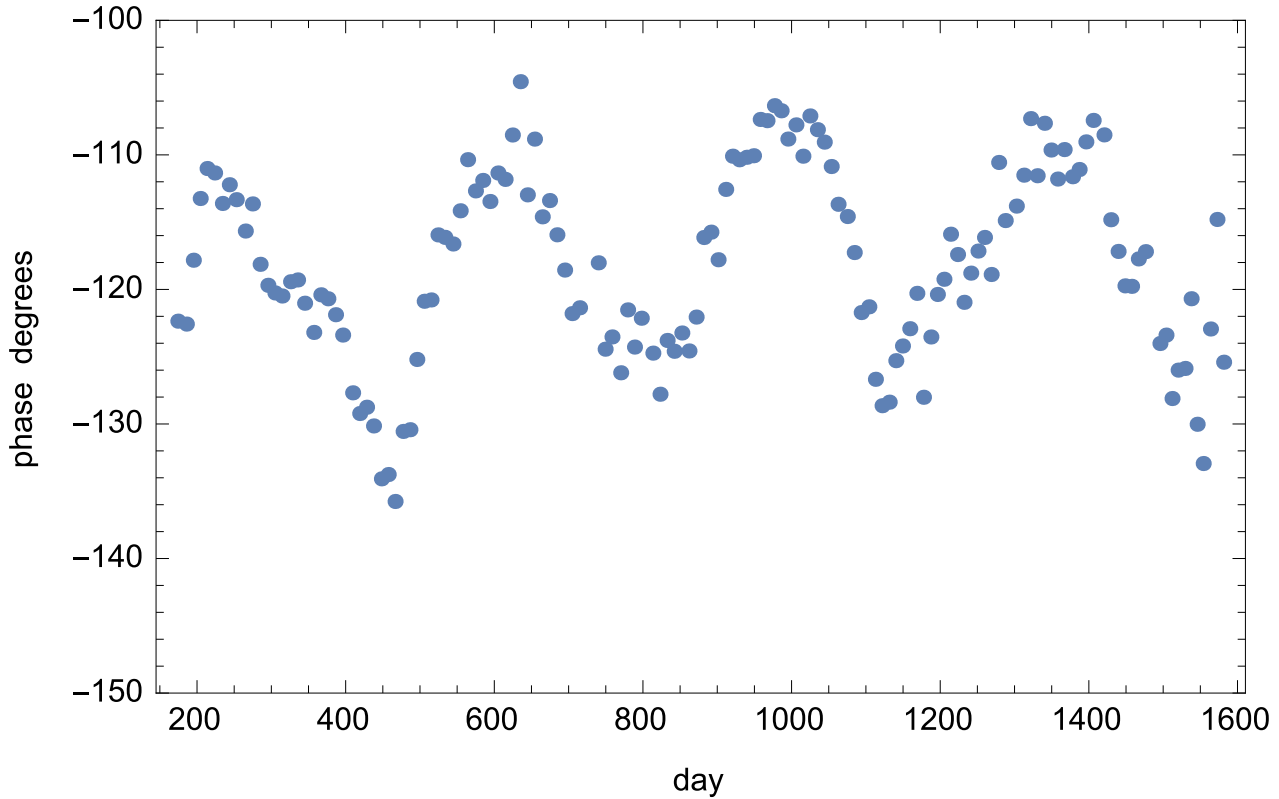}
\hspace{2pc}
\caption{Phases of the main oscillation with period 0.54634 d computed for 9 days-long segments of the PCA-cleaned light curve. The zero-point
of phase determinations is day 1000.}
\label{phall.fig}
\end{figure}

We divide each quarter into intervals of 9 days long (typically, 10 intervals per quarter) and compute a Lomb's periodogram fit with
a fixed period of $P=0.54634$ d, clipping the flux to $\pm 30$ ppt around the median value to remove the largest flares. The phase of this
dominating oscillation was then computed for each interval as $\arctan(a_c, a_s)$, where $a_c$ and $a_s$ are the coefficients of the fit
for the model functions $\cos(2 \pi t'/P)$ and $\sin(2 \pi t'/P)$, respectively. The times $t'$ are shifted to a common zero-point approximately
in the middle of the mission, i.e., $t'=t-1000$, which helps to de-correlate the phase and frequency for the entire collection of observations.
Fig. \ref{phall.fig} shows segmented phase estimates in degrees across the entire mission. We find surprisingly large variations
which look systematic and periodic. It resembles the strictly periodic signals in ETV data for eclipsing stars which are caused by the Doppler
effect or the light travel time effect (sometimes inaccurately called R{\o}mer effect) due to the orbital motion of the target around a
third body, i.e., in triple systems \citep[see, e.g.,][]{con}. In our case, whatever the source of the Doppler delay is, a stable single-frequency
flux signal (irrespective of its origin) can be phase-modulated by a barycentric radial velocity variation. However, a full-scale periodogram
analysis of the same segments of data with period as a free parameter does not show any periodic variation, being quite flat in time with a mean
of 0.54642, median 0.54638, and standard deviation 0.00065 d. This discrepancy is puzzling because a physical Doppler effect should change
both the frequency and phase, since the former is the time derivative of the latter. Certainly, the sensitivity of the phase determinations
is much greater because the amplitude of modulation gets multiplied by its frequency in the corresponding period modulation, which is a small
number. From the half-amplitude and frequency of modulation in Fig. \ref{phall.fig}, the main mode period is expected to be modulated by
roughly 0.4\%, or 0.00021 d. This is indeed difficult to detect given the random error of period determinations. Hence, a Doppler-like modulation
may be present in the segmented period estimates but we do not see it.

\begin{figure}[htbp]
  \centering
  
  \includegraphics[width=5in]{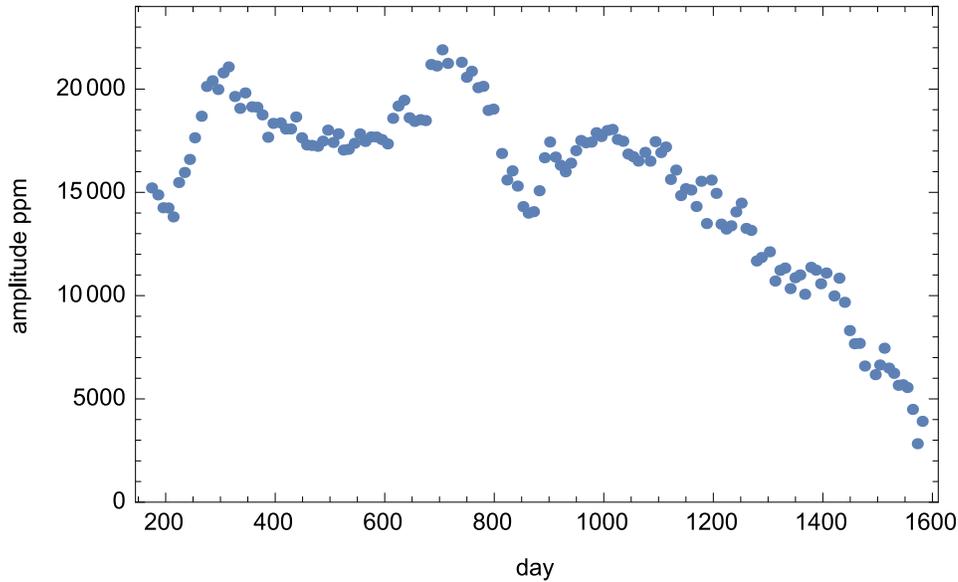}
  \caption{Amplitude of the dominating mode of flux oscillation with periods around 0.5464 d.}
  \label{amp.fig}
\end{figure}

The amplitude of oscillations significantly varies with time too --
see Fig. \ref{amp.fig}. The amplitude is computed as the RSS of the estimated coefficients for the main mode,
$a=\sqrt{a_c^2+a_s^2}$. There is no obvious periodicity, but the amplitude strongly declines
toward the end of the mission. In addition to the oscillating pattern the light curve also features numerous
flares with amplitudes greatly exceeding the oscillation amplitude. This makes the least-squares model of flux vulnerable to perturbations
and potentially dependent on many additional parameters. To avoid fitting too many parameters we are not interested in, we tried to use a model 
where the amplitude is not present at all. 

For each quarter, we determine the median flux $f_m$ and set an auxilliary variable $z$:

\begin{eqnarray}
  \label{eq:2}
  z=0 \ \hbox{if} f<f_m \nonumber\\
z=1 \ \hbox{if} f\ge f_m \nonumber
\end{eqnarray}

We then fit a logistical model:
\begin{eqnarray}
  \label{eq:3}
  p(z=1) = \frac{1}{1+e^{-x}} \nonumber\\
  p(z=0) = \frac{e^{-x}} {1+e^{-x}} \nonumber
\end{eqnarray}

where
\begin{eqnarray}
  \label{eq:4}
  x=s \cos\left(  \Omega_{osc} t + \phi  \right) \\
\phi = \phi_0+ A_{\phi} cos (\Omega_{\phi} t + \phi_{\phi})
\end{eqnarray}
with $s$ being a scale parameter, $\Omega$ the frequency, and
$\phi$ the phase of the high-frequency oscillation. $A_{\phi} $ is the amplitude of phase variation, $\Omega_{\phi}$
is the phase variation frequency. This model is simplified in the sense that it includes only a single harmonic of the phase
variation rather than a Fourier series, and the main harmonic of the flux oscillation, but it will prove adequate in the
following analysis.

\begin{figure}[htbp]
\centering
    
\includegraphics[width=6in]{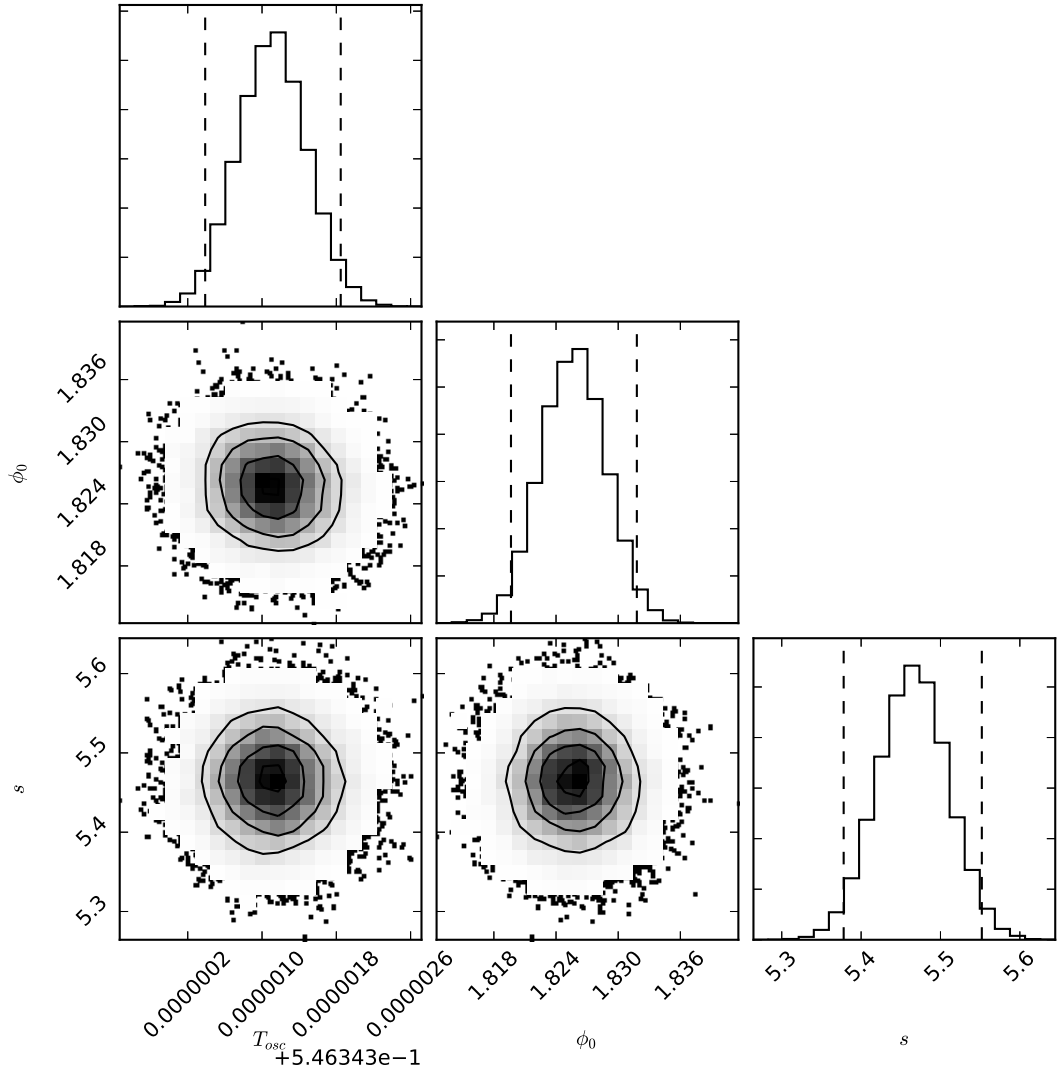}
\caption{Projections of log-likelihood distribution for flux oscillation period $T_{osc}$, flux oscillation phase and scale $s$.}
\label{logl1.fig}\end{figure}

\begin{figure}[htbp]
  \centering

\includegraphics[width=6in]{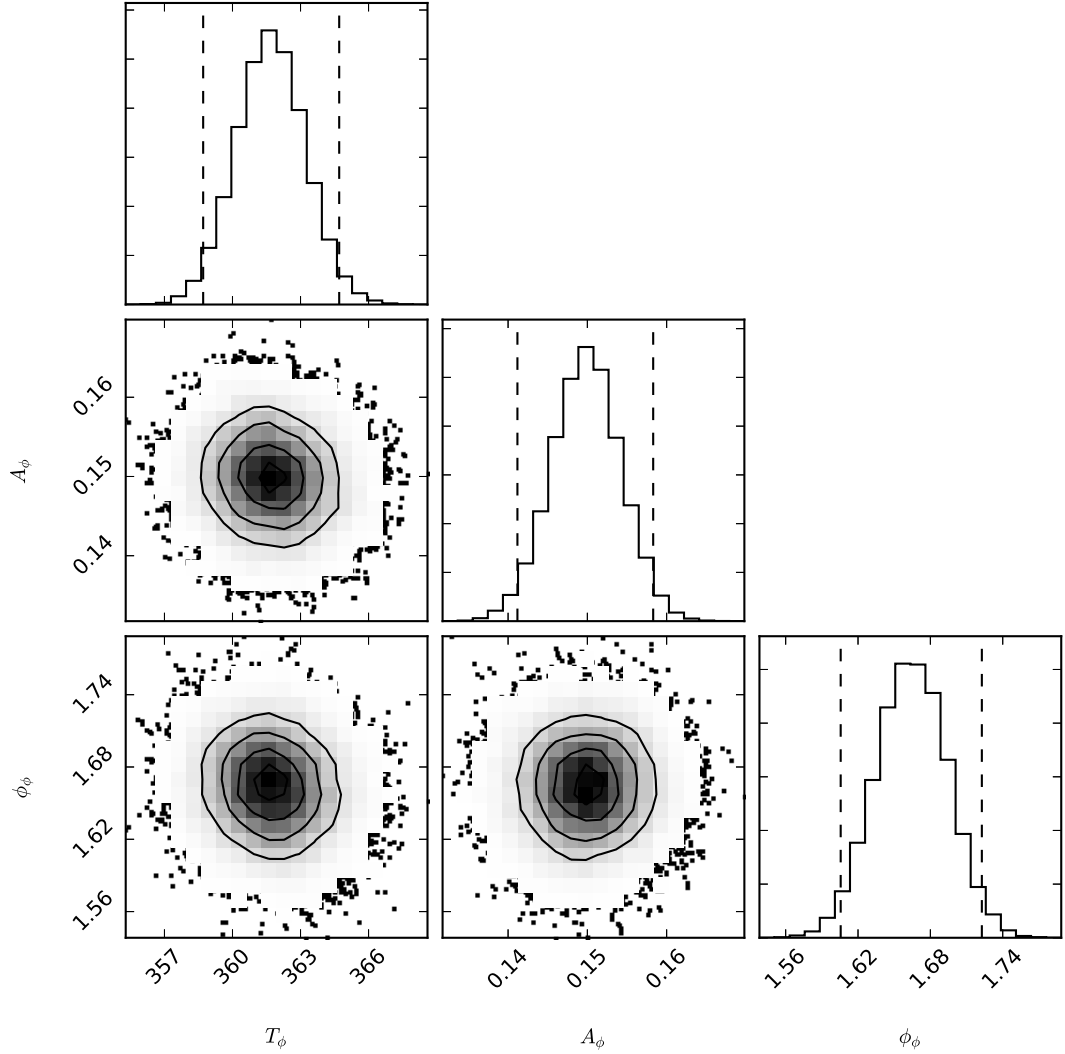}
  \caption{Projections of log-likelihood distribution for phase period $T_0$, phase amplitude $A_\phi$, and phase $\phi_0$.\label{logl2.fig}}

\end{figure}

This likelihood parametrization has certain advantages. There is no
need to fit a time-dependent phase amplitude. We also avoid fitting the possibly variable shape of flux oscillations and a special
treatment of numerous flares and upward spikes in the light curve. 
The log-likelihood function
\begin{equation}
  \label{eq:6}
  \log  L\left( s, \Omega,  \phi_0 ,  A_{\phi}  , \Omega_{\phi} ,
    \phi_{\phi}, \right)  = \sum_i \log p(z_i)
\end{equation}
is sampled by the
Markov Chain Monte-Carlo package EmCee  \citep{for}. The resulting density plots of the log-likelihood function of the free parameters in
Eq. \ref{eq:4} using 160000 samples after discarding initial 100000 burn-in samples are shown in Figs. \ref{logl1.fig}, \ref{logl2.fig}. The
fit is well-defined with this model with tight and round-shaped confidence areas. The fitting parameters and their 95\%
confidence intervals are summarized in Table \ref{tab:fitted}.

\begin{table}[htbp]
  \centering
\bgroup
\def\arraystretch{1.5}
  \begin{tabular}{| c | c  l |}
    \hline
\hline

  $T_{osc}$ & 	0.5463441 &$^{+7.3\cdot 10^{-7}}_{-7.2\cdot
              10^{-7}}$\\

$\phi_0$ &	1.8257& $^{	+0.0061}_{-0.006} $\\

$s$& 	5.466& $^{+0.008}_{-0.009}$\\

$T_{\phi} $ &	361.68&$^{+3.02}_{-2.97} $\\

$A_{\phi}$	& 0.150&$^{+0.008}_{-0.009}$\\

$\phi_{\phi}$ & 1.664&$^{+0.06} _{-0.06}$\\

\hline

  \end{tabular}
\egroup

  \caption{Fitted parameters}
  \label{tab:fitted}
\end{table}

\section{Interpretation}
\label{int.sec}

Periodic variations of flux are not infrequent at the sensitivity level of the Kepler photometric data. They can be caused by a number of
physical phenomena, but generally, the basic underlying mechanisms are related to 1) intrinsic pulsations; 2) rotation and nonuniform surface
brightness distribution; 3) binarity effects. Even though pulsations of different excitation origin were theoretically predicted, they
proved to be too small for a positive detection with Kepler even in dedicated short-cadence observations \citep{rod}. Besides, the
predicted periods of pulsations are considerably shorter than our main oscillation mode ($0.5463441(7)$ d). In the following discussion,
we will omit this possibility due to the null result of previous studies.

The remaining two causes, binarity and rotation/activity, may often be interrelated. The rate of rotation and magnetic activity
of main sequence dwarfs slows down with age due to rotation braking \citep{sku, bar}, but this process can be arrested and reversed in tight
binary stars by the angular momentum transfer via tidal energy dissipation \citep[e.g.,][]{egg}. As a result, binaries with orbital
periods below a certain threshold (a few days, depending on the spectral type and evolutionary status) are practically all synchronously
rotating \citep{lev}. The characteristic time of spin-down for single M dwarfs is $\sim400$ Myr \citep{sta}. If a single M dwarf is older than
the age of the Hyades, it is expected to rotate with a longer period than the observed value for \st{}.

Binary M-dwarfs, on the other hand, can acquire and maintain fast rotation rates for billions of years. The tidal evolution of the orbit
is directed toward gradual circularization and further tightening of the orbit. The tidal friction, however, drops to zero for a completely
circularized and synchronized binary of zero obliquity \citep{hut}. Such tidal equilibrium pairs remain fast-rotating and, therefore,
magnetically active for their lifetimes, because the reservoir of orbital kinetic energy is still large. How can an originally wider pair
reach this state without the assistance of significant tidal dissipation? \citet{kis} proposed a mechanism where a tertiary companion,
if significantly inclined with respect to the inner orbit generates secular oscillations of the inner orbit's eccentricity and inclination
and boosts the tidal dissipation in the inner pair. This mechanism finds support in the observation that most active binaries known to us
often reside in hierarchical multiple systems \citep{make}. The emerging interpretation of \st{} consistent with this scenario is a
hierarchical triple system with a mid-type M dwarf as the optically brightest component whose fast rotation is driven by a non-eclipsing
secondary, and a sufficiently massive, more distant tertiary component.

It would be tempting to explain the quasi-periodic variations of phase we see in Fig. \ref{phall.fig} as the direct consequence of the
orbital motion of the inner binary about the tertiary via the light travel time effect (LTTE). Using formulae from \citet{mur},
adjusted to direct phase measurements, we performed a linear optimization fitting for the following Keplerian LTTE model:
\eb
\phi(t)=\frac{4\pi a_0 \sin i}{c T_{osc}}\left[ \cos\omega\frac{\sqrt{1-e^2}}{e}\sum_{k=1}^{\infty}\frac{J_k(ke)\sin\,knt}{k} +
\sin\omega\sum_{k=1}^{\infty}\frac{J_{k-1}(ke)-J_{k+1}(ke)}{2k} \cos\,knt\right]
\ee
where $a_0$ is the semimajor axis of the observed oscillating star, $c$ is the speed of light, $i$ is the inclination of the
orbit to the line of sight, $\omega$ is the argument of pericenter, $e$ is the eccenticity, $n$ is the mean motion ($n=2\pi/P_{\rm orb}$),
and $J_k$ is the Bessel function of the first kind. The fitting parameters in this model are $a_0\sin i$, $\omega$, and $e$. From the former,
we can estimate the lower bound mass of the outer companion if we assume a certain mass of the inner binary (or single oscillator) from
Kepler's third law. Our best fit provides an $a_0\sin i = 117(6)\times 10^6$ km and an eccentricity consistent with 0 but uncertain to 0.1.
Assuming a mass of 0.2 $M_{\sun}$ for the target star, the mass of the hypothetical 362-day orbiter is 0.77 $M_{\sun}$. Since this
companion is fainter in the optical, it could be a white dwarf or a neutron star.

There is significant evidence putting this interpretation of the phase variations in doubt. The curve in Fig. \ref{phall.fig} appears
more complex than the simple sinusoid our Keplerian fit has produced. There may be other signals of non-cyclic nature present. The 0.546-day
oscillation is not perfectly sinusoidal either, which is seen in the light curve (Fig. \ref{singflare.fig}) and is confirmed by periodograms
where we find conspicuous second and third harmonics with periods $T_{osc}/2$ and $T_{osc}/3$. Irrespective of the nature of this non-sinusoidal
oscillation, an LTTE effect should modulate the frequency of these harmonics too, resulting in a similar cyclic variation of their phase.
We performed a similar segmented phase determination for the second harmonic of oscillation and the result (not shown in this paper for brevity)
indicates large-amplitude smooth variation that is quite dissimilar to the sinusoid with a 362-day period. This means that not only the
phase and amplitude of the main oscillation wave vary, but also its shape. 

These peculiar properties are easier explained in the framework of a large brightness feature on the rotating surface, such as a giant spot.
Can stellar spots live for longer than 4 years? While most sunspots exist only for days to weeks, active RS CVn-type binaries were found
to have long-lived, giant spots with an average lifetime of 2.2 yr, stretching to 4.5 yr \citep{str94,str09}. Generally, tight binaries
sport longer-lived spots than quiescent single stars. A star can have a polar cap-like large spot living for decades if its meridional
flow is faster than the solar flow \citep{hol}. \citet{tra} suggested a persistent near-pole hot spot on the surface on a differentially rotating
inclined star to explain the anti-correlated eclipse timing variations observed with Kepler for contact binaries. Differential rotation
with the equatorial region rotating systematically faster than the polar regions, is believed to be the direct consequence of the Coriolis
force acting on convective turbulence \citep{kit}, so it can be common for M stars. The secular decrease of the amplitude of the main oscillation
mode in Fig. \ref{amp.fig} can be explained as regular decay of the giant spot toward the end of the main mission (although the exact
physical processes involved there are not well understood even for sunspots). The different behavior of the second oscillation harmonic
whose pronounced amplitude and phase variations are not correlated with the main mode, can be reasonably expected if the shape and brightness
distribution changed in the course of 4 years of observation. 

The main challenge to this hypothesis is the cyclic behavior of the phase of main mode oscillation (Fig. \ref{phall.fig}). Generally,
the phase variation may be caused by a persistent active region or a spot migrating or drifting on the surface in longitude, or,
in combination with a differential rotation, in latitude. Both directed drifts will effectively change the observed frequency of
flux variations. The open issue is what kind of process can make this migration be quasi-periodic in time. The period of variation
362(3) d is perhaps too short to be related to a possible intrinsic magnetic activity cycle. The character of main mode phase variation suggests 
a periodic component of the surface rotation rate, i.e., libration. If the high rate of rotation is maintained by a close low-mass
companion, and there is a more massive tertiary companion on a much wider, inclined orbit, the libration of the pericenter argument
of the inner orbit prescribed by the Lidov-Kozai mechanism  \citep{egg,kis,kir} can cause such periodic rotational acceleration. 
Even in the framework
of the spot hypothesis, a hierarchical triple system remains most suitable to explain the peculiar properties of \st. 

\section{Summary}
We consider it unlikely that \st{} is very young such as a T Tauri star or a YSO. The closest young stars to the Sun are identified
in loose moving groups and usually have wide companions with approximately the same space velocity and age. No such moving group members
have been suggested in the Kepler field. Aside of the unusual infrared colors, the star is likely to be a powerful and variable
source of X-ray radiation with giant X-ray flares. We also detect super-flares in the optical Kepler band, with the largest
event on day 1443 releasing $\sim 4\times 10^{34}$ erg of energy within 3.5 hours. These properties put the mid-type M dwarf in the category
of extremely active UV Cet-type stars. The energy released during the largest flare is similar to the event detected by \citet{ost}
for DG CVn in the optical bandpass.

Our astrometric VIM analysis reveals that each flux variation is rather closely mirrored in the detected centroid position (Fig. \ref{singflare.fig}). 
This indicates the presence of blended sources in the digital apertures of the target star. The dominating VIM effect probably comes from
a rather distant, partially blended source (companion 2 in Fig. \ref{map.fig}), but a more detailed view suggests at least one
additional source which seems to have a significant azimuthal motion with respect to the target (Fig. \ref{vimpa.fig}). Thus, the first clue
of binarity or multiplicity of \st{} comes from Kepler astrometry. Unfortunately, the amplitudes and shapes of the most significant VIM
effects are altered by what we suspect to be a manifestation of the ``brighter-fatter" effect, which also puts in question the directional
information.

The short-period quasi-sinusoidal flux periodicity can be traced across the entire mission span, but its amplitude and phase underwent
large secular changes. As the phase shows large-amplitude quasi-periodical variations, the amplitude appears to be fairly constant
during the first half of the mission span, then declining from peak values of 20 ppt to less than 5 ppt (Fig. \ref{amp.fig}). Could
the time-variable component in VIM direction and the oscillation amplitude behavior come from the same physical process? In principle,
the quasi-period phase modulation with a period of 362(3) d can be explained as a Doppler shift (LTTE) due to the orbital acceleration
around a sufficiently massive (at least, $0.77\,M_{\sun}$) companion, but the amplitude decay does not seem to be consistent with this interpretation.
A more plausible scenario involves a giant spot or an activity region on the photosphere of the fast-rotating star. This would also account for
the dissimilar changes in the phase and amplitude of the second harmonic of the 0.54634 d oscillation. The finding at odds with this
interpretation is the cycling character of the phase variations. A more sophisticated model may be required to explain the scope
of properties, such as a hierarchical triple system with a tight inner binary (but not eclipsing) and a more distant tertiary causing
the pericenter argument to librate around $90\degr$ or $270\degr$ via the Lidov-Kozai mechanism \citep{egg}. 
We propose such a system as our best
educated guess, because it also explains the fast rotation of the M dwarf and its high degree of activity. The tidal dissipation of
kinetic energy due to a periodic boost to the inner pair's eccentricity causes the orbit shrink, increasing the rate of rotation
of the tidally synchronized components. The model is consistent with the highest levels X-ray and flaring activity observed in
close BY Dra and RS CVn systems.

What could be the nature of the hypothetical companion to \st? A matching ultraviolet source is found in the deep GALEX survey of the Kepler
field by \citet{olm}. The measured near-UV magnitude ($N_{UV}$) is 20.412(44) mag. Assuming a distance of 21 pc, the absolute near-UV magnitude
is $M_{\rm NUV}=18.8$ mag. With the accurate Gaia DR1 magnitude $G=12.633$, the visual color $N_{UV} - G=7.78$ mag which places the star
far to the left (too blue) of the well-defined main sequence constructed for nearby dwarfs (Makarov, 2017, in preparation). The expected
color at this absolute magnitude is 10.04 mag. The difference is too large to be associated with the elevated level of magnetic activity
observed in a small fraction of nearby M dwarfs and is more likely to indicate the presence of a hot degenerate companion. The absolute
$N_{UV}$ magnitude for such photometrically blended M--WD pairs is close to that of the WD component. This allows us to tentatively estimate
the spectral type at D6, i.e., a relatively cool and evolved WD. \citet{mor} find that at late spectral types (M4--M6), WD+dM pairs flare 50\% as frequently as field dMs.

The ultimate judgement on whether a solar-mass companion exists in the system will be delivered by accurate radial velocity measurements.
\st{} was part of a spectroscopic survey conducted by \citet{tera, terb}, but unfortunately, only one determination was obtained for this
object (R. Terrien 2016, priv. comm.). A follow-up K2 photometric series would also be useful to determine if the short-period oscillation
of flux has died out, or, otherwise, if its phase is still consistent with the value observed during the main Kepler mission. 
\label{sum.sec}
\section*{Acknowledgments}
This paper includes data collected by the Kepler mission. Funding for the Kepler mission is provided by the NASA Science Mission directorate.
Some of the data presented in this paper were obtained from the Mikulski Archive for Space Telescopes (MAST). STScI is operated by the Association of Universities for Research in Astronomy, Inc., under NASA contract NAS5-26555. Support for MAST for non-HST data is provided by the NASA Office of Space Science via grant NNX09AF08G and by other grants and contracts. This research has made use of ``Aladin sky atlas" developed at CDS, Strasbourg Observatory, France. This work has made use of data from the 
European Space Agency (ESA)
mission {\it Gaia} (\url{http://www.cosmos.esa.int/gaia}), processed by
the {\it Gaia} Data Processing and Analysis Consortium (DPAC,
\url{http://www.cosmos.esa.int/web/gaia/dpac/consortium}). Funding
for the DPAC has been provided by national institutions, in particular
the institutions participating in the {\it Gaia} Multilateral Agreement.
This research has made use of the VizieR catalogue access tool, CDS,
 Strasbourg, France. The original description of the VizieR service was
 published in A\&AS 143, 23

\label{lastpage}

\end{document}